\documentclass{article}

\usepackage{arxiv}

\usepackage[utf8]{inputenc} 
\usepackage[T1]{fontenc}    
\usepackage{hyperref}       
\usepackage{url}            
\usepackage{booktabs}       
\usepackage{amsfonts}       
\usepackage{nicefrac}       
\usepackage{microtype}      
\usepackage{graphicx}
\usepackage{lipsum}

\title{Future ALICE upgrades for Run 4 and Beyond}

\author{
  H. Sebastian Scheid\\
  for the ALICE Collaboration\\
  \texttt{s.scheid@cern.ch} \\
}

\begin{document}

\maketitle
\begin{abstract}
ALICE pursues several upgrades to further extend its physics reach. 
For Run 4, ALICE is pioneering the construction of truly cylindrical tracking layers, which will improve the measurements of heavy-flavour hadrons and dielectrons.
In addition, a Forward Calorimeter (FoCal) for the measurement of direct photons is being developed to access the gluon distributions of nucleons and nuclei at low $x$.
For Run 5 and beyond, ALICE 3 is proposed. It combines a unique high-resolution vertex detector with a silicon pixel tracker and modern particle identification solutions over a large acceptance.
This will permit heavy-flavour and dielectron measurements with unprecedented precision to access fundamental properties and dynamics of the quark-gluon plasma (QGP). 
We will discuss the upgrade plans, report on R\&D results for ITS3 and FoCal, and present the requirements and concepts for ALICE 3.

\end{abstract}
  
\section{Introduction}
To access the dynamics of the QGP and earlier phases of heavy-ion collisions it is necessary to measure probes that pierce the veil of hadron production at the phase boundary of hadronisation.
Here two major topics are heavy-flavour (HF) and electromagnetic (EM) probes. The former, produced in the earliest hard parton-parton interactions, evolve with the medium and carry information about the collision dynamics. These imprint in the correlation of HF hadrons or the production of multi-charm baryons. EM probes, produced in all stages of the collision, carry their information to the detector unscathed. The spectra, flow or polarisation of virtual photons can shed light on the early temperature and the evolution of the medium.
Furthermore, measurements of direct photons, at forward rapidity are proposed to provide unique access to the low-$x$ gluon structure of protons and nuclei. These will be possible with the FoCal, combining a high-resolution electromagnetic Si-W calorimeter with a conventional hadronic calorimeter~\cite{ALICECollaboration:2719928}.
Furthermore,  to replace the three innermost layers of its inner tracking system (ITS),  ALICE is pioneering the usage of bent, wafer-scale pixel sensors~\cite{Musa:2703140}. This makes it possible to construct truly cylindrical layers with close to no support structures, reducing the material budget significantly. The resulting improvement in pointing resolution will allow better measurements of heavy-flavour hadrons and dielectrons. 
Taking these efforts to a new level the proposed ALICE 3 experiment will combine a unique vertex tracker, with tracking and particle identification based on cylindrical barrel layers combined with forward disks to cover a large rapidity together with low momentum acceptance~\cite{ALICE:2803563}. 

\section{The ALICE Upgrades for Run 4}
Installation of the FoCal and the replacement of the inner three layers of the ALICE ITS  are foreseen during long shutdown 3 (LS3) from 2026 to 2028.

\subsection*{A forward calorimeter}
The FoCal will consist of a high-granularity Si-W sampling sandwich calorimeter to measure EM showers (FoCal-E), followed by a metal-scintillator sampling calorimeter to measure the associated hadrons. The detector will be installed at $3.4 < \eta < 5.8$, 7~m away from the nominal interaction point (IP), opposite of the muon arm.
The planned setup for the Focal-E consists of two digital high-granularity (HG) layers with a pixel pitch of $30 \times 30$ $\mathrm{\mu m^{2}}$ and 18 layers analogue read out pads with a $1\times1$~$\mathrm{cm^{2}}$ pitch. These are housed in a stack separated by 3.5~mm W-absorbers, corresponding to $\approx 1 X_{0}$ each. The HG layers are planned to be placed at position 5 and 10 in the stack to provide a separation of two close-by EM showers.
This makes possible to measure direct photons and neutral mesons together with associated hadrons at forward rapidities, enabling the study of the gluon structure of hadronic matter down to $x \approx 10^{-6}$ (and small $Q^{2}$). Here possible non-linear effects of QCD such as gluon saturation could emerge.

A prototype equipped with HG layers in the absorber stack was tested at the CERN SPS in 2021 with beams consisting of different particles at energies of 20/40/60/80~GeV. 
\begin{figure}[htb]
    \centerline{
        \includegraphics
        [width=4.5cm]
        {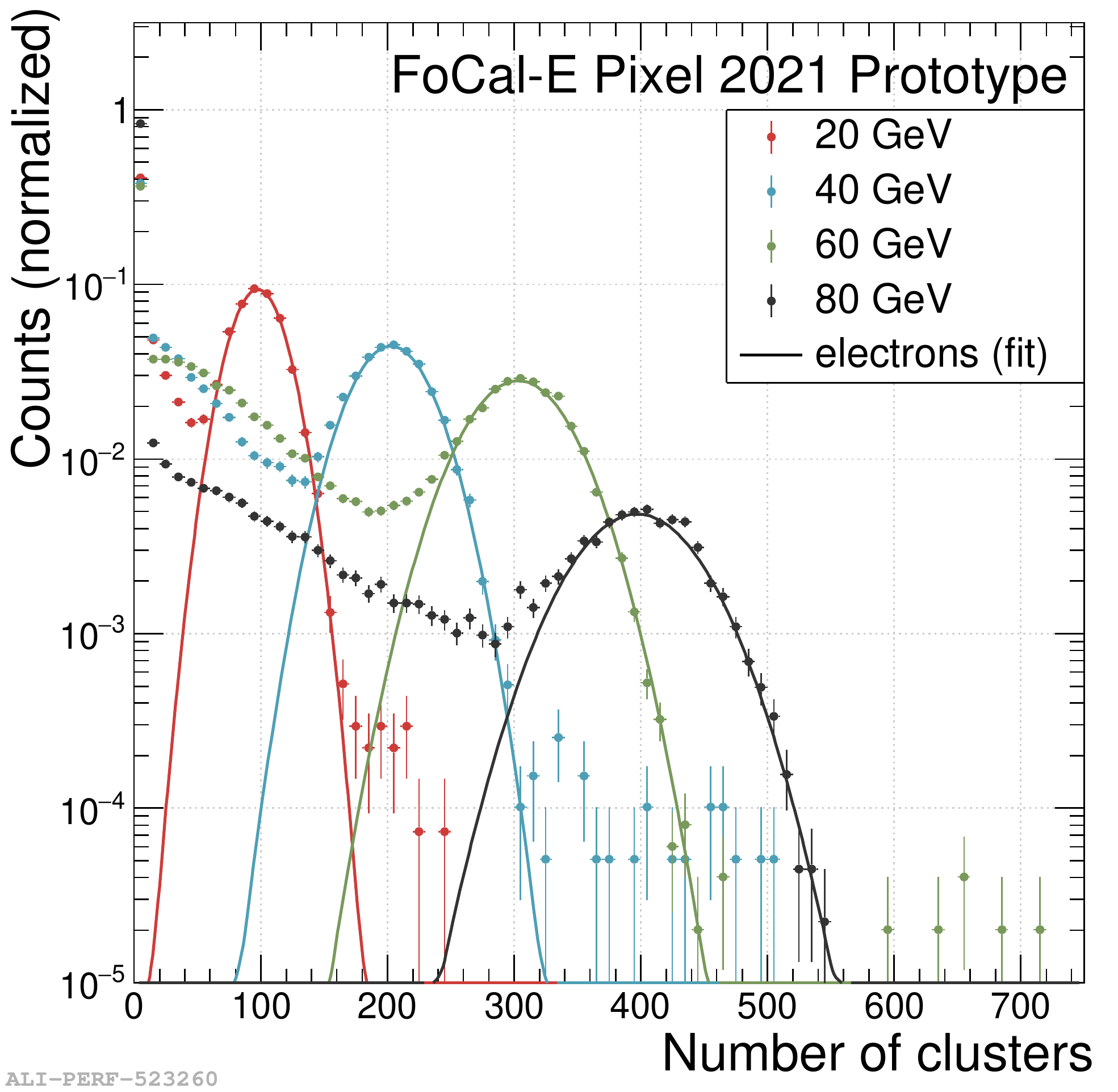}
        \includegraphics
        [width=8.5cm,trim={3cm 0 21cm 1.5cm},clip]
        {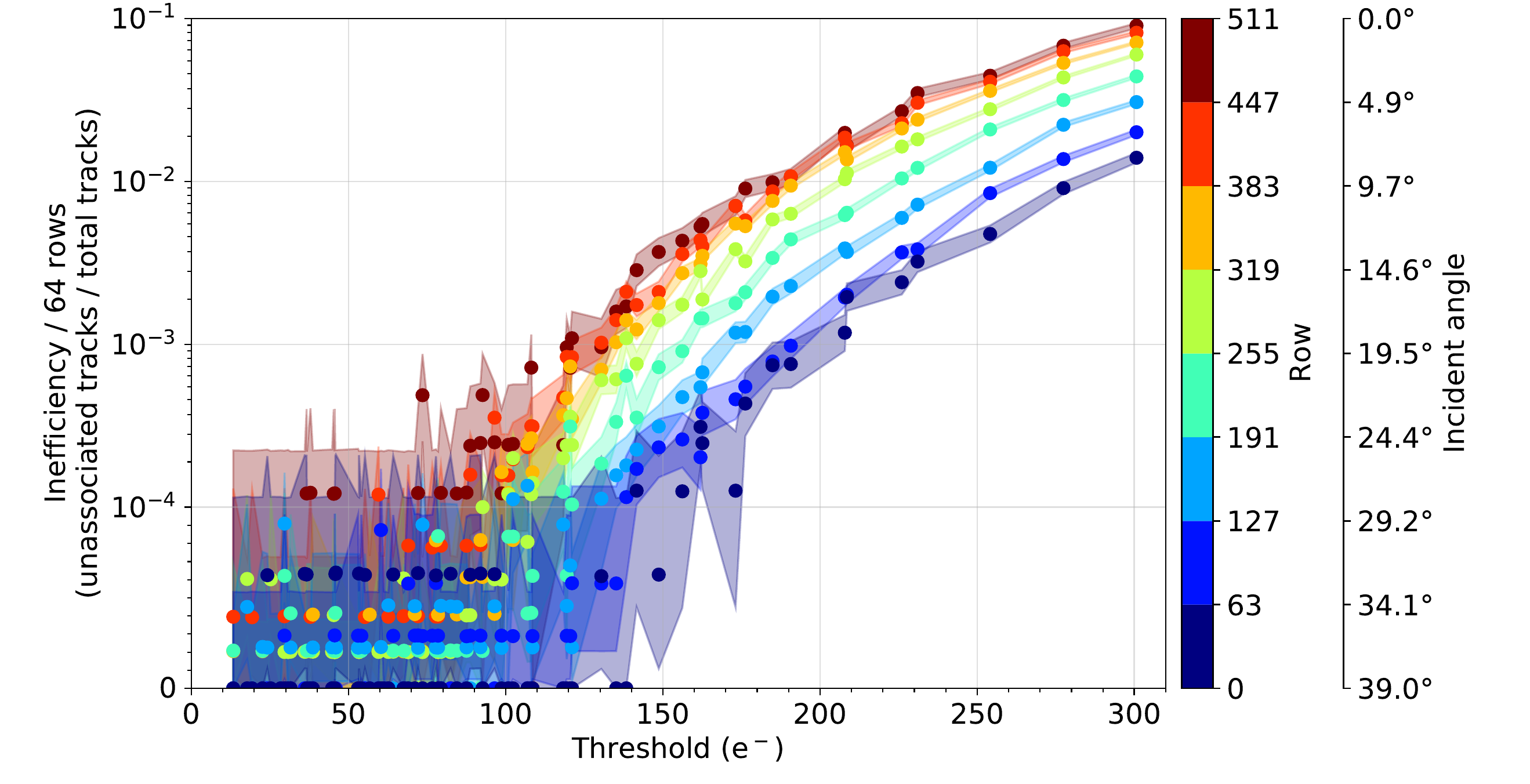}
    }
    \caption{Left: Normalised counts as a function of the number of clusters measured with in the second pixel layer at different beam energies. The gaussian parameterisation of the EM component is shown as a solid line.
    Right: Inefficiency of a bent ALPIDE chip as a function of threshold. 
    The different colored lines correspond to the incident angle of the incoming beam~\cite{ALICEITSproject:2021kcd}.}
    \label{Fig:Focal}
\end{figure}
The results for the number of clusters in the $2^{\rm nd}$ layer is shown in Fig.~\ref{Fig:Focal} (left). The position of the electron peak was determined with a parameterisation combining a gaussian with a  background component (not shown). The change in the overall shape of the spectra as a function of energy indicates a change in the particle chemistry of the beam depending on the energy. The electron peak position is well described by a detailed simulation.
While a linear dependence of the number of clusters on the beam energy was observed in layer 10 (after $10X_{0}$), layer 5 ($5X_{0}$) shows saturation effects. This indicates that the electromagnetic shower is not fully developed at the position of the measurement yet. 

\subsection*{A truly cylindrical tracker}
The ITS3 project aims to replace the three innermost layers of the current ITS with truly cylindrical ones, with a material budget reduced to the bare minimum, i.e. the active silicon. This can be achieved by integrating power and data routing in the chip and lowering the power consumption so that cooling by air is sufficient. Using sensor stitching, the chips can be produced in the size of silicon wafers. Thinned to only a few tens of $\mathrm{\mu m}$ the material becomes flexible enough to be bend,  which allows the reduction of the support structure. Within the R\&D efforts large-scale mechanical tests were done with dummy wafers as well as functional tests with thinned ALPIDE chips.


It was demonstrated that 50 $\mathrm{\mu m}$ silicon dummys can be bent to the desired 18/24/30~mm curvature radii and fixated with carbon foam ribs.
Together with a new thinner and smaller beam-pipe this allows the placement of the first detection layer closer to the IP (from 22 to 18 mm). 
The combination of material reduction (factor 6) and the layers placed in closer proximity to the IP enhances the pointing resolution by a factor 3.
This for example improves the significance of the measurement of $\Lambda_{c} \rightarrow pK^{-}\pi^{+}$ by a factor 4$-$5~\cite{Musa:2703140}.
Figure~\ref{Fig:Focal} (right) shows that the inefficiency as a function of pixel threshold of an ALPIDE chip bent to a radius of $22\pm1$~mm showed no deterioration in its performance at the nominal working point of about 100 $e^{-}$~\cite{ALICEITSproject:2021kcd}.
The construction of the ITS3 relies on the production of chips in the 65~nm process to reach 30~cm wafer scale chips, via sensor stitching.
Within the same test setup as the bent ALPIDE, the first digital pixel test structures produced in a 65~nm process showed to be at least 99\% efficient.

\section{ALICE 3 - A Next-Generation Heavy-Ion Experiment}
At the time of RUN4 conclusion (2032), there will still be open questions concerning the physics of the QGP. 
Two critical areas that were identified to contribute to answer them are multi-differential precision measurement of dileptons and a systematic measurement of (multi-)heavy-flavour hadrons. Where the dilepton spectra are sensitive to the time-dependence of the early stage temperature, the HF measurements probe the thermalisation in, as well as hadronisation from the medium.
At the LHC both measurements are highly correlated, since the HF contribution is the main background to the dilepton measurements. A detector with unprecedented tracking capabilities will be necessary to perform such measurements at the LHC.

The ALICE 3 detector concept extends current R\&D within the ITS3 project to a unique device for the measurement of heavy-ion collisions.
The vertex detector will be located in a secondary vacuum within the beam-pipe and as close as 5 mm from the IP, the aperture of the LHC beam at top energies. For safe operation the detector can be retracted at beam injection.
Together with further silicon-based tracking layers a pointing resolution of 20 $\mathrm{\mu m}$ at $p_{\rm T} = 100$~MeV/$c$ can be achieved.
Particle identification will be possible over a broad momentum range thanks to Time-of-Flight stations at 19 and 85 cm accompanied by a Ring-Imaging-Cherenkov detector based on an aerogel radiator.
To ensure particle identification also at large rapidities, 405 cm away from the interaction point, along the beam direction, a TOF station is foreseen together with a RICH. 
Continuous coverage of the PID via RICH is achieved by adjusting the refractive indices for the barrel (1.03) and the forward direction (1.006).
The tracking and particle identification detectors, together with a electromagnetic calorimeter will be housed in a superconducting magnet system. Operated at 2~T, a momentum resolution of $~1\%$ at $\eta = 0$ can be achieved. A muon system consisting of an absorber and tracking stations will be installed outside the magnet.
For the measurement of photons at smallest $p_{\rm T}$ a dedicated conversion tracker at forward rapidities is foreseen. Installation of the detector is planned during LS4 in 2033 and 2034.

The unprecedented tracking and pointing capabilities of the proposed detector, together with its large rapidity coverage and low $p_{\rm T}$ reach, will introduce a new era of measurements in the heavy-flavour and dilepton sector.
Performance estimates show that for the first time it will be possible to measure the azimuthal correlation of $\mathrm D^{0}-\bar{\mathrm D}^{0}$ pairs in Pb$-$Pb collisions. At low $p_{\rm T}$ this type of measurement is directly sensitive to the decorrelation of HF quarks in the QGP~\cite{Nahrgang:2013saa}.
Furthermore, the tracking layers at very small radii give the possibility for a new technique in the measurement of strange hadrons, i.e. to directly track them together with their decay products. This method of {\it strangeness tracking} will significantly boost the precision of measurements in the sector of multi-charm hadrons, since the prompt and non-prompt strange hadrons can be separated. This will give access to measurements of particles like the $\Omega_{\rm cc}$, which decays with an $\Omega$ in its decay chain. Due to the early production of heavy-flavour quarks, the measurement of multi-charm baryons will give unique access to thermalisation and hadronisation mechanisms from the QGP.

\begin{figure}[htb]
    \centerline{
    \includegraphics[width=12.5cm]{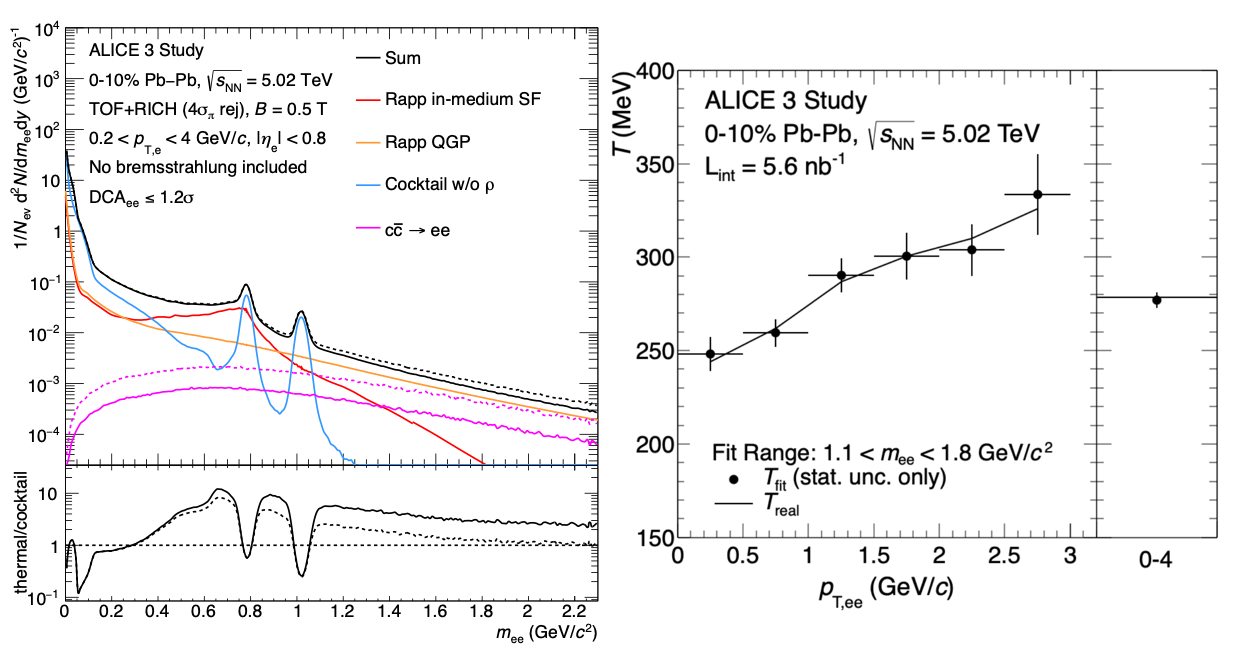}
    }
    \caption{Left: Expected dielectron yields in central Pb$-$Pb events after different selection criteria are applied (top) and the ratio for thermal over hadronic contributions (bottom). Solid lines indicate the projections for ALICE 3, dashed lines for ITS3.
    Right: Medium temperature extracted from the inverse slope parameter in momentum intervals in simulations (points) and model calculations (line).}
    \label{Fig:Dielectrons}
\end{figure}

The excellent detection capabilities of heavy-flavour decays will also benefit the measurements of thermal radiation from the QGP in the form of dielectrons. 
Here the partonic contribution to the thermal radiation is expected to be detectable at masses above 1 GeV/$c^{2}$, where at LHC energies the dielectrons from semi-leptonic decays of HF hadrons dominate. Using the pair distance of closest approach of the dielectron to the primary vertex ($\rm DCA_{\rm ee}$) it is possible to reject 96 (98)\% of the charm (beauty) contribution, while conserving 76\% of the prompt thermal contribution. The expected sources and their abundances after selecting only pairs with $\rm DCA_{\rm ee} < 1.2 \sigma$ are shown in Fig.~\ref{Fig:Dielectrons} (left). 
The superior pointing resolution with respect to ITS3 improves the rejection of the HF and the corresponding uncertainty at masses above 1 GeV/$c^{2}$, which is the sensitive region for a measurement of the QGP temperature. In Fig.~\ref{Fig:Dielectrons} (right) the expected values of the temperature determined from the slope of the emission spectrum based on calculations including the full time evolution of the medium~\cite{Rapp:1999ej} are shown for different selections of the transverse momentum of the pair. The statistical uncertainties reflect the expectation for one month of data taking, while systematic uncertainties are not shown. A measurement like this is in particular interesting since the production of dielectrons with larger momenta is favoured at higher energy densities, i.e. earlier in the evolution of the QGP. 
This measurement allows to access the time dependence of the emission of thermal radiation. The expected statistical precision allows also to measure the elliptic flow or the polarisation of thermal dielectrons.

\section{Conclusion}
ALICE has made significant progress to extend the physics program of the LHC for Run 4 with the operation of bent silicon sensors and the demonstration of the FoCal concept in test beams. Furthermore, the submission of the ALICE 3 proposal to the LHCC and their recommendation to proceed with the project constitute important milestones in the LHC heavy ion program.


\bibliographystyle{plain}
\bibliography{literature}

\end{document}